\documentclass[10pt,aps,pre,twocolumn,showpacs,showkeys,superscriptaddress,byrevtex]{revtex4}
\usepackage{amssymb}
\usepackage{graphics}
\begin{document}
\hyphenation{ma-cro-state}
\hyphenation{mi-cro-state}
\hyphenation{mi-cro-ca-no-ni-cal}
\hyphenation{ca-no-ni-cal}

\title{An Introduction to the Thermodynamic and
Macrostate Levels of\\ Nonequivalent Ensembles%
\footnote{Contribution to the Proceedings of the Second Sardinian
International Conference on News and Expectations in Thermostatistics
(NEXT 2003), Villasimius (Cagliari), Sardegna, Italy.}
}

\author{H. Touchette}

\email{htouchet@alum.mit.edu}

\address{\mbox{Department of Physics and School of Computer Science, McGill University, 
Montr\'eal, Qu\'ebec, Canada H3A 2A7}}

\author{R. S. Ellis}

\email{rsellis@math.umass.edu}

\affiliation{Department of Mathematics and Statistics, University of Massachusetts,
Amherst, Massachusetts 01003 USA}

\author{B. Turkington}

\affiliation{Department of Mathematics and Statistics, University of Massachusetts,
Amherst, Massachusetts 01003 USA}

\date{\today}

\begin{abstract}

This short paper presents a nontechnical introduction to the problem of
nonequivalent microcanonical and canonical ensembles. Both the thermodynamic
and the macrostate levels of definition of nonequivalent ensembles are
introduced. The many relationships that exist between these two levels are
also explained in simple physical terms.

\end{abstract}



\maketitle

\section{Introduction}

Most textbooks of statistical mechanics (see, e.g., \cite
{reif1965,huang1987,balian1991,landau1991,salinas2001}) have sections
devoted to demonstrating that the microcanonical and canonical
ensembles---the two sets of equations used to calculate the equilibrium
properties of many-body systems---always give the same predictions. The
arguments given are most often not actual proofs, but variations of an
argument originally put forward by Gibbs in his seminal treatise \cite
{gibbs1902} claiming that the canonical ensemble should be equivalent to the
microcanonical ensemble in the thermodynamic limit. Gibbs's reasoning
basically is that although a system having a fixed temperature does not
have, theoretically speaking, only one definite value of energy (the
canonical distribution is spread over many energy values), the fluctuations
of the system's energy should become negligible in comparison with its total
energy in the limit where the volume of the system tends to infinity. In
this limit, the thermodynamic limit, the system should thus appear to
observation as having a definite value of energy---the very hypothesis which
the microcanonical ensemble is based on \footnote{
``For the average square of the anomalies of the energy, we find an
expression which vanishes in comparison to the square of the average energy,
when the number of degrees of freedom is indefinitely increased. An ensemble
of systems in which the number of degrees of freedom is of the same order of
magnitude as the number of molecules in the bodies with which we experiment,
if distributed canonically, would therefore appear to human observation as
an ensemble of systems in which all have the same energy.'' \cite[p. xi]
{gibbs1902}}. The conclusion then apparently follows, namely:\ both the
microcanonical and the canonical ensembles should predict the same
equilibrium properties of many-body systems in the thermodynamic limit of
these systems independently of their nature.

Gibbs's treatise is a milestone in the development of equilibrium
statistical mechanics. Hence, it is not surprising that it has had a great
influence on advancing the idea that it does not matter whether the
equilibrium properties of a system are calculated from the point of view of
the microcanonical or the canonical ensemble; i.e., whether they are
calculated as a function of the energy or the temperature of the system,
respectively. Gibbs himself found an explicit formula expressing the
temperature of the perfect gas as an invertible function of its internal
energy per particle, thus showing that the perfect gas has the same
microcanonical and canonical equilibrium properties. Later on, many other
many-body systems were shown to behave similarly. Faced with such evidence,
it seems then logical to argue, as most physicists now do, that the
equilibrium, energy-dependent properties of any large enough system can
always be related in a one-to-one fashion with its temperature-dependent
properties. But the problem, unfortunately, is that this is not always the
case.

Since the 1960's, many researchers, starting with Lynden-Bell and Wood \cite
{lynden1968}, have found examples of statistical mechanical models
characterized at equilibrium by microcanonical properties which have no
equivalent within the framework of the canonical ensemble. The
nonequivalence of the two ensembles has been observed for these models both
at the thermodynamic and the macrostate levels of description of statistical
mechanics, and, recently, a complete mathematical theory of nonequivalence
of ensembles due to Ellis, Haven and Turkington \cite{ellis2000} has
appeared in an effort to better understand this phenomenon. Our goal in this
short paper is to offer a simplified and nontechnical presentation of this
theory and to emphasize its physical interpretation so as to give an
accessible explanation of the phenomenon of nonequivalent ensembles.

We shall start in the next section by explaining first how the
microcanonical and canonical ensembles can be nonequivalent at the
thermodynamic level, which is the level that has been studied the most so
far. In Section \ref{sMacro}, we then discuss a more fundamental definition
of nonequivalent ensembles introduced in \cite{ellis2000} by Ellis, Haven
and Turkington---the \textit{macrostate} level of nonequivalence of
ensembles---and explain intuitively how this level is related to the
thermodynamic level of nonequivalent ensembles. We conclude by providing in
Section \ref{sConc} a list of references which illustrate many of the
results mentioned here, and offer some thoughts about the possibility of
experimentally observing nonequivalent ensembles.

\section{Thermodynamic Nonequivalence of Ensembles}

\label{sThermo}

In the physics literature, there exist two basic ways by which the
microcanonical and canonical ensembles have come to be defined as being
nonequivalent at the level of the thermodynamic quantities of a system 
\footnote{
See the concluding section for references.}. The first way is global in
essence. It consists in making a statement about the overall shape of the
microcanonical entropy function, which for a system consisting of $n$
particles, is commonly defined by the limit 
\begin{equation}
s(u)=\lim_{n\rightarrow \infty }\frac 1n\ln \Omega (u),
\end{equation}
where $\Omega (u)$ denotes the density of microstates of the system having a
mean energy $u$ \cite{reif1965}. From the global point of view, then, we
have thermodynamic nonequivalence of ensembles whenever the graph of $s$
contains one or more nonconcave dips that make the first derivative of $s$ a
non-monotonic function of $u$.

Such a definition is likely to appear odd for physicists because most of
them were taught to think that the microcanonical entropy $s$ is an always
concave function of $u$ \footnote{
A misconception propagated, once again, by most textbooks on statistical
mechanics; see \cite{wannier1966,thirring2002} for notable exceptions.}. But
the truth is that this function \textit{can} be nonconcave, as many
researchers have pointed out in recent years, and the dramatic consequence
of this insight is the following. If the function $s(u)$ is not concave on
its entire domain of definition, then this function cannot be expressed as
the Legendre transform, or the Legendre-Fenchel transform \footnote{
The Legendre-Fenchel transform is a generalization of the common Legendre
transform which can be applied to non-differentiable functions, and which
reduces to the Legendre transform when applied to differentiable functions;
see \cite{rockafellar1970,ellis1985}.}, of the free energy function, the
basic thermodynamic function of the canonical ensemble defined as 
\begin{equation}
\varphi (\beta )=\lim_{n\rightarrow \infty }-\frac 1n\ln Z(\beta ),
\end{equation}
where $Z(\beta )$ denotes the partition function at inverse temperature $
\beta $ \cite{reif1965}. In fact, only in the case where $s^{\prime}$
exists for all $u$ and $s^{\prime}$ is known to be monotonic in $u$ does $s$
equal the Legendre transform of $\varphi (\beta )$; in symbols, 
\begin{equation}
s(u)=\beta (u)u-\varphi (\beta (u)),  \label{l1}
\end{equation}
where $\beta (u)=s^{\prime }(u)$ \footnote{
One can superficially understand from (\ref{l1}) why a nonconcave $s(u)$
cannot be expressed as the Legendre transform of $\varphi (\beta )$. If $
s^{\prime }(u)$ is non-monotonic in $u$, then the differential equation $
\beta =s^{\prime }(u)$, which is the basis of the Legendre transform, will
have multiple root solutions in $u$ for some values of $\beta $. In such a
case, the Legendre transform cannot induce an invertible mapping of the
values of $u$ to the values of $s^{\prime }$ as usually done when $s^{\prime
}$ is a monotonic function of $u$.}.

\begin{figure*}[tb] 
\centering
\resizebox{0.7\textwidth}{!}{\includegraphics{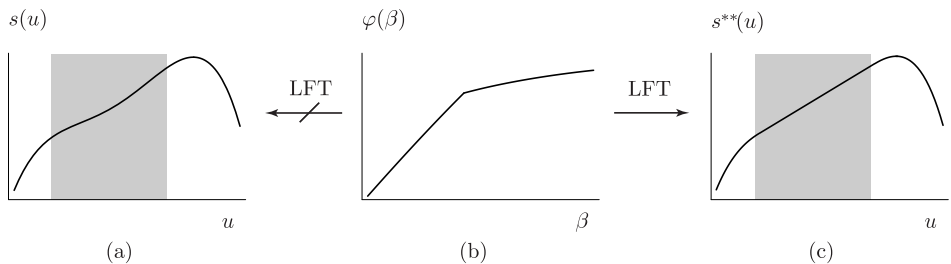}}
\caption{(a) Nonconcave microcanonical entropy function $s(u)$ and (c) its
concave envelope $s^{**}(u)$. The latter function is derived as the
Legendre-Fenchel transform (LFT) of the free energy function $\varphi(\beta)$.
The concave points of $s(u)$ are defined as the mean energy values $u$ 
for which we have
$s(u)=s^{**}(u)$. The nonconcave points of $s(u)$, on the other hand,
correspond to the values of $u$ for which $s(u)\neq s^{**}(u)$ (see shaded region).}
\label{maxwellfem4}
\end{figure*}

At this point, we turn to the second thermodynamic way of defining
equivalent and nonequivalent ensembles. What we would like to have now is a
local criterion---as opposed to the global criterion just presented---for
deciding when the microcanonical and the canonical ensembles are equivalent
or nonequivalent. To define such a criterion, we reverse in a way the logic
of the global definition by directly defining the double dual of $s(u)$ by
the Legendre-Fenchel transform of the free energy $\varphi (\beta )$ as
follows: 
\begin{equation}
s^{**}(u)=\inf_\beta \{\beta u-\varphi (\beta )\}.  \label{lf2}
\end{equation}
Such a function can be shown to be concave on its domain of definition, in
addition to be equal to the minimal concave function majorizing $s(u)$ for
all $u$ \cite{rockafellar1970}. Given this property of $s^{**}(u)$, it is
thus to be expected that if the graph of $s$ possesses any nonconcave dips,
then there will be points of $s$ where $s^{**}(u)\neq s(u)$. This
observation is, in effect, what enables us to give a local definition of
equivalent and nonequivalent ensembles. Namely, if $s^{**}(u)=s(u)$, then
the microcanonical and the canonical ensembles are said to be \textit{
thermodynamically equivalent} at the mean energy value $u$. In such a case,
the two ensembles are equivalent precisely in the sense that the value of $s$
at $u$ can be calculated from the point of view of the canonical ensemble by
taking the Legendre-Fenchel transform of $\varphi (\beta )$, as in (\ref{lf2}).
Conversely, we say that the two ensembles are \textit{thermodynamically
nonequivalent} at the mean energy value $u$ whenever $s^{**}(u)\neq s(u)$,
i.e., whenever 
\begin{equation}
s(u)\neq \inf_\beta \{\beta u-\varphi (\beta )\}.
\end{equation}
An illustration of these definitions is given in Figure \ref{maxwellfem4}.

\section{Macrostate Nonequivalence of Ensembles}

\label{sMacro}

The macrostate-level definition of nonequivalent ensembles has the same
relationship to the thermodynamic level definition of nonequivalent
ensembles as statistical mechanics has to thermodynamics: it is a deeper and
hence more refined level of conceptualization from which the other level can
be derived. A virtue of the macrostate point of view is also that
nonequivalent ensembles are defined in a more natural way mathematically
than their thermodynamic counterparts. Choosing a macrostate, say $m$,
related to the statistical mechanical model of interest, one needs indeed
only to proceed as follows:

\begin{enumerate}
\item  Calculate the set $\mathcal{E}^u$ of equilibrium values for $m$ in
the microcanonical ensemble with mean energy $u$.

\item  Calculate the set $\mathcal{E}_\beta $ of equilibrium values for $m$
in the canonical ensemble with inverse temperature $\beta $.

\item  Compare $\mathcal{E}^u$ and $\mathcal{E}_\beta $ for the different
values of $u$ and $\beta $.
\end{enumerate}

If all the members of the microcanonical set $\mathcal{E}^u$ can be put in a
one-to-one correspondence with all the members of a canonical set $\mathcal{E
}_\beta $, i.e., if there exists $\beta $ such that $\mathcal{E}^u=\mathcal{E
}_\beta $, then macrostate equivalence of ensembles is said to hold. On the
other hand, if there exists a set $\mathcal{E}^u$ (resp., $\mathcal{E}_\beta 
$) having at least one member which cannot be found in any set $\mathcal{E}
_\beta $ for all $\beta $ (resp., any set $\mathcal{E}^u$ for all $u$), then
macrostate nonequivalence of ensembles is said to hold. Hence, we have
macrostate nonequivalence of the microcanonical and canonical ensembles if
one of two ensembles is richer than the other.

These definitions of macrostate equivalence and nonequivalence of ensembles
are natural and require us only to check the possible relationships that may
exist between the sets $\mathcal{E}^u$ and $\mathcal{E}_\beta $. But as
definitions they are also useless because they only classify. They predict
nothing. What we would like to have, of course, is a list of simple
criteria---based, for example, on the knowledge of thermodynamic quantities
like $s(u)$ or $\varphi (\beta )$---to decide whether or not the
microcanonical and the canonical ensembles are equivalent at the level of
macrostates. Could there be, for instance, any connections between the
thermodynamic level of equivalence or nonequivalence of ensembles and the
macrostate level of equivalence or nonequivalence of ensembles which could
enable us to say anything about the latter level?

In answer to this question, Ellis, Haven and Turkington have provided in 
\cite{ellis2000} a number of rigorous mathematical results which classify
all possible relationships that can exist between $\mathcal{E}^u$ and $
\mathcal{E}_\beta $ for all $u$ and $\beta $ based on knowledge of the
microcanonical entropy function $s(u)$. Their most important results about
these relationships are summarized in the following items 1, 2 and 3 
\footnote{
For simplicity, the differentiability of $s(u)$ is assumed throughout.}.

\begin{enumerate}
\item  (Full equivalence of ensembles). If $s(u)=s^{**}(u)$ at $u$, and $s(u)
$ is not locally flat around $u$, then $\mathcal{E}^u=\mathcal{E}_\beta $
for $\beta =s^{\prime }(u)$.

\item  (Nonequivalence of ensembles). If $s(u)\neq s^{**}(u)$ at $u$, then $
\mathcal{E}^u\cap \mathcal{E}_\beta =\emptyset $ for all $\beta $. Thus,
thermodynamic nonequivalence of ensembles at $u$ implies nonequivalence of
ensembles at the macrostate level. This also shows that the microcanonical
ensemble can be richer than the canonical ensemble.

\item  (Partial equivalence of ensembles). If $s(u)=s^{**}(u)$ at $u$, but $
s(u)$ is locally flat around $u$, then $\mathcal{E}^u\subsetneq \mathcal{E}
_\beta $ for $\beta =s^{\prime }(u)$; i.e., $\mathcal{E}^u$ is a proper
subset of $\mathcal{E}_\beta $ in this case. From this item and the first
one, we conclude that thermodynamic equivalence of ensembles at $u$ implies
either full equivalence of partial equivalence of ensembles at the level of
macrostates.
\end{enumerate}

It is not our intention in this paper to prove these mathematical results;
full proofs can be found in \cite{ellis2000}. What we would like to do,
however, is to attach a physical meaning to these results so as to give the
reader an intuitive feeling for their validity. To that end, we shall simply
make use of Gibbs's argument that was stated at the beginning of this paper,
although we shall take care now of specifying which values of the mean
energy are realized at equilibrium in the canonical ensemble, and---this is
the crucial point---how many of them are realized.

First, let us note that the values of the mean energy which are realized at
equilibrium in the canonical ensemble must correspond to the global
minimizers of the function $I_\beta (u)=\beta u-s(u)$. This is a well-known
fact of equilibrium statistical mechanics; see \cite{reif1965}. From this
result it is then not difficult to verify that if $s(v)=s^{**}(v)$ and if $s$
is not locally flat around $v$, then $I_\beta (u)$ has a unique global
minima located at $v$ for $\beta =s^{\prime }(v)$ \cite{touchette32003}. In
such a case, Gibbs's reasoning is thus true; namely, in the limit where $
n\rightarrow \infty $, the canonical ensemble with inverse temperature $
\beta =s^{\prime }(v)$ does indeed reduce to the microcanonical ensemble
with mean energy $v$ because, in this very limit, the canonical ensemble
assumes the unique equilibrium mean-energy value $v$. This intuitively leads
us to the result stated in item 1.

In the case of a point $v$ where $s^{**}(v)\neq s(v)$, we can work out a
similar argument; however, the result that we have to use now is the
following. If $s^{**}(v)\neq s(v)$ at $v$, then the mean energy value $v$
can never be realized at equilibrium in the canonical ensemble for all $
\beta $ \cite{touchette32003}. In other words, the canonical ensemble must
jump over all values of the mean energy for which we have thermodynamic
nonequivalence of ensembles. In this case, we intuitively expect to have $
\mathcal{E}^u\cap \mathcal{E}_\beta =\emptyset $ for all $\beta $, as stated
in item 2. Note, as an aside, that this argument leads us to an interesting
result: any region of thermodynamic or macrostate nonequivalence of
ensembles must give rise to a first-order canonical phase transition 
\footnote{
See \cite{ellis2000,touchette2003,touchette42003} for a proof of this result
based on the properties of the Legendre-Fenchel transform.}.

The final result that we must discuss to complete our interpretation of the
macrostate level of nonequivalent ensembles is the result in item 3 about
partial equivalence. For this result, one can verify that if $s(v)=s^{**}(v)$
at $v$, but $s$ is locally flat around $v$, then the canonical ensemble at
inverse temperature $\beta =s^{\prime }(v)$ gives rise to multiple
equilibrium values of the mean energy; specifically all $v$ such that $\beta
=s^{\prime }(v)$. In this situation, we accordingly expect to see the
canonical ensemble reduce not to a single microcanonical ensemble, but to
many coexisting microcanonical ensembles, each one corresponding to a
mean-energy value realized at equilibrium in the canonical ensemble. This
represents, of course, nothing but the emergence of a state of coexisting
phases which normally takes place at first-order phase transitions. In more
symbolic notations, we thus expect to have 
\begin{equation}
\mathcal{E}_\beta =\mathcal{E}^{v_1}\cup \mathcal{E}^{v_2}\cup \cdots ,
\end{equation}
where $v_1,v_2,\ldots $, denote all canonical equilibrium values of the mean
energy satisfying $s^{\prime }(v_i)=\beta $. Under the assumption that $
\mathcal{E}^{v_i}\neq \emptyset $ for all $i=1,2,\ldots $, we then recover
the statement of item 3, namely $\mathcal{E}^v\subsetneq \mathcal{E}_\beta $
with $\beta =s^{\prime }(v)$.

\section{Concluding Remarks}

\label{sConc}

The global approach to the problem of thermodynamic nonequivalence of
ensembles based on the observation of nonconcave dips in the graph of $s(u)$
has been suggested by a number of people. We already mentioned the work of
Lynden-Bell and Wood \cite{lynden1968}, who seem to have been the first to
observe such dips in the entropy function of certain gravitational many-body
systems (see \cite{lynden1999} for a historical account). Similar
observations have also been reported in the same context by Thirring and
Hertel \cite{thirring1970,hertel1971}, and by Gross \cite
{gross1997,gross2001} more recently. For a recent survey of the subject, the
reader is invited to consult the comprehensive collection of papers edited
by Dauxois et al.~\cite{draw2002}; 
it covers a wide range of physical models for which
nonconcave anomalies of the microcanonical entropy function have been
observed, and contains much information about the physics of these models.
Other examples of applications related to lattice-spin systems can be found
in \cite
{kiessling1997,dauxois2000,ispolatov2000,antoni2002,borges2002,barre2001,barre22002}
.

The second approach to the thermodynamic nonequivalence of ensembles
presented here, which explicitly focuses on the properties of
Legendre-Fenchel transforms and on the local properties of $s(u)$, is due
for the most part to Ellis, Haven and Turkington \cite{ellis2000}, and Eyink
and Spohn \cite{eyink1993} (see also \cite{thirring2002}). The works of
these authors represent also the primary sources of information for the
theory of macrostate nonequivalence of ensembles. Various illustrations of
this theory, dealing with statistical models of turbulence, can be found in 
\cite{ellis2002,ellis42002}. We mention finally our recent work \cite
{touchette2003} on the mean-field Blume-Emery-Griffiths spin model which can
be consulted as an easily accessible introduction to the material surveyed
in this paper.

To conclude, we would like to call attention to the fact that no physical
experiment has been designed to explicitly measure a discrepancy between the
microcanonical and canonical equilibrium macrostate properties of a system.
In an attempt to enter this \textit{terra incognita}, there is perhaps no
better way to start than to explore the deep connection that exists between
nonequivalent ensembles and first-order canonical phase transitions \cite
{gross1997,gross2001,touchette2003}. Thus, one can look for a system which
displays such a type of phase transitions, and then try to imagine a way to
block the transition so as to be able to vary the system's energy at will
within the range of energy values skipped by the canonical ensemble. The
energy of the system, as such, need not be frozen indefinitely in time in
order for that system to be microcanonical. In practice, what is required is
to be able to select any value of the energy, and to make sure that the
relaxation time of the energy fluctuations of the system is greater than the
observation time (high Deborah-number-system \cite{reiner1964}). If one is
to verify the theory of macrostate nonequivalence of ensembles, then it is
within the range of canonically-forbidden energy values---i.e., within the
range of the latent heat---that microcanonical nonequivalent values of
macrostates are to be found and nowhere else.

\section*{Acknowledgments}

One of us (H.T.) would like to thank the organizing committee of the NEXT
2003 Conference for partially financing his participation in this
conference, as well as the Mathematics and Statistics Department at the
University of Massachusetts at Amherst for providing a serene environment in
which this paper could be written. The research of R.S.E.\ and B.T.\ was
supported by grants from the National Science Foundation (NSF-DMS-0202309
and NSF-DMS-0207064, respectively). The research of H.T.\ was supported by
FCAR (Qu\'{e}bec) and the Cryptography and Quantum Information Laboratory of
the School of Computer Science at McGill University.

\bibliography{next2003htouc}

\end{document}